\documentclass[11pt,twoside]{article}
\usepackage{./asp2014}

\aspSuppressVolSlug
\resetcounters

\begin{document}

\title{Spectropolarimetric Observations of an Arch Filament System with GREGOR}
\author{H. Balthasar,$^1$ P. G\"om\"ory,$^2$ S.~J.~Gonz\'alez~Manrique,$^{1,3}$ 
C. Kuckein,$^1$ A.~Ku{\v c}era,$^2$ P. Schwartz,$^2$ T.~Berkefeld,$^4$ 
M.~Collados,$^5$ C.~Denker,$^1$ A. Feller,$^6$ A. Hofmann,$^1$ D.~Schmidt,$^7$ 
W.~Schmidt,$^4$ M.~Sobotka,$^8$ S.~K.~Solanki,$^{6,9}$ D.~Soltau,$^4$ 
J.~Staude,$^1$ K.~G.~Strassmeier,$^1$ and O.~von~der~L\"uhe$^4$}
\affil{$^1$Leibniz-Institut f\"ur Astrophysik Potsdam (AIP), 14482 Potsdam, Germany; \email{hbalthasar@aip.de}}
\affil{$^2$Astronomical Institute of the Slovak Academy of Sciences, Tatransk\'a Lomnica, Slovak Republic}
\affil{$^3$Universit\"at Potsdam, Institut f\"ur Physik und Astrophysik, 14476 Potsdam, Germany}
\affil{$^4$Kiepenheuer-Institut f\"ur Sonnenphysik, 79104 Freiburg, Germany}
\affil{$^5$Instituto de Astrof{\'\i}sica de Canarias, 38205 La Laguna (Tenerife), Spain}
\affil{$^6$Max-Planck-Institut f\"ur Sonnensystemforschung, 37077 G\"ottingen, Germany}
\affil{$^7$National Solar Observatory, Sunspot NM 88349, U.S.A.}
\affil{$^8$Astronomical Institute, Academy of Sciences of the Czech Republic, Ond{\v r}ejov, Czech Republic}
\affil{$^9$Kyung Hee University, Yongin, Gyeonggi-Do, 446 701, Republic of Korea}

\paperauthor{H. Balthasar}{hbalthasar@aip.de}{}{Leibniz-Institut f\"ur Astrophysik Potsdam (AIP)}{Physics of the Sun}{Potsdam}{}{D-14482}{Germany}
\paperauthor{P. G\"om\"ory}{gomory@ta3.sk}{}{Astronomical Institute of the Slovak Academy of Sciences}{}
{Tatransk\'a Lomnica}{}{}{Slovakia}
\paperauthor{S.J. Gonz\'alez Manrique}{smanrique@aip.de}{}{Leibniz-Institut f\"ur Astrophysik Potsdam (AIP)}{Physics of the Sun}{Potsdam}{}{D-14482}{Germany}
\paperauthor{C. Kuckein}{ckuckein@aip.de}{}{Leibniz-Institut f\"ur Astrophysik Potsdam (AIP)}{Physics of the Sun}{Potsdam}{}{D-14482}{Germany}
\paperauthor{A. Ku{\v c}era}{akucera@ta3.sk}{}{Astronomical Institute of the Slovak Academy of Sciences}{}
{Tatransk\'a Lomnica}{}{}{Slovakia}
\paperauthor{P. Schwartz}{pschwartz@ta3.sk}{}{Astronomical Institute of the Slovak Academy of Sciences}{}
{Tatransk\'a Lomnica}{}{}{Slovakia}
\paperauthor{T. Berkefeld}{berkefeld@leibniz-kis.de}{}{Kiepenheuer-Institut f\"ur Sonnenphysik}{}{Freiburg}{}{D_79104}{Germany}
\paperauthor{M. Collados}{cv@iac.es}{}{Instituto de Astrof{\'\i}sica de Canarias}{}
{La Laguna}{Tenerife, Islas Canarias}{E-38205}{Spain}
\paperauthor{C. Denker}{cdenker@aip.de}{}{Leibniz-Institut f\"ur Astrophysik Potsdam (AIP)}{Physics of the Sun}{Potsdam}{}{D-14482}{Germany}
\paperauthor{A. Feller}{feller@mps.mpg.de}{}{Max-Planck-Institut f\"ur Sonnensystemforschung}{}{G\"ottingen}{}{D-37077}{Germany}
\paperauthor{A. Hofmann}{ahofmann@aip.de}{}{Leibniz-Institut f\"ur Astrophysik Potsdam (AIP)}{Physics of the Sun}{Potsdam}{}{D-14482}{Germany}
\paperauthor{R. Schlichenmaier}{schliche@leibniz-kis.de}{}{Kiepenheuer-Institut f\"ur Sonnenphysik}{}{Freiburg}{}{D_79104}{Germany}
\paperauthor{D. Schmidt}{dschmidt@nso.edu}{}{National Solar Observatory}{Sacramento Peak}{Sunspot}{NM}{88349}{U.S.A.}
\paperauthor{W. Schmidt}{wolfgang@leibniz-kis.de}{}{Kiepenheuer-Institut f\"ur Sonnenphysik}{}{Freiburg}{}{D_79104}{Germany}
\paperauthor{M. Sigwarth}{msig@leibniz-kis.de}{}{Kiepenheuer-Institut f\"ur Sonnenphysik}{}{Freiburg}{}{D_79104}{Germany}
\paperauthor{M. Sobotka}{michal.sobotka@asu.cas.cz}{}{Astronomical Institute of the Academy of Sciences of the Czech Republic}{}
{Ond{\v r}ejov}{}{}{Czech Republic}
\paperauthor{S.K. Solanki}{solanki@mps.mpg.de}{}{Max-Planck-Institut f\"ur Sonnensystemforschung}{}{G\"ottingen}{}{D-37077}{Germany}
\paperauthor{D. Soltau}{soltau@leibniz-kis.de}{}{Kiepenheuer-Institut f\"ur Sonnenphysik}{}{Freiburg}{}{D-79104}{Germany}
\paperauthor{J. Staude}{jstaude@aip.de}{}{Leibniz-Institut f\"ur Astrophysik Potsdam (AIP)}{Physics of the Sun}{Potsdam}{}{D-14482}{Germany}
\paperauthor{K.~G Strassmeier}{kstrassmeier@aip.de}{}{Leibniz-Institut f\"ur Astrophysik Potsdam (AIP)}{}{Potsdam}{}{D-14482}{Germany}
\paperauthor{O. von der L\"uhe}{ovdluhe@leibniz-kis.de}{}{Kiepenheuer-Institut f\"ur Sonnenphysik}{}{Freiburg}{}{D_79104}{Germany}

\begin{abstract}
      We observed an arch filament system (AFS) in a sunspot group with the 
      GREGOR Infrared Spectrograph attached to the GREGOR solar telescope. 
      The AFS was located between the leading sunspot of negative polarity 
      and several pores of positive polarity forming the following part of the 
	  sunspot group.  
      We recorded five spectro-polarimetric scans of this region. The spectral 
	  range included the spectral lines Si~{\sc i} 1082.7~nm, He~{\sc i} 
	  1083.0~nm, and Ca~{\sc i} 1083.9~nm. 
      In this work we concentrate on the silicon line which is 
      formed in the upper photosphere. The line profiles are inverted 
      with the code `Stokes Inversion based on Response functions' to obtain 
	  the magnetic field vector. The line-of-sight velocities are determined 
	  independently with a Fourier phase method. Maximum velocities are found
	  close to the ends of AFS fibrils.
      These maximum values amount to 2.4~km~s$^{-1}$ next to the pores and to 
	  4~km~s$^{-1}$ at the sunspot side. 
      Between the following pores, we encounter an area of negative polarity 
	  that is decreasing during the five scans. We interpret this by new 
	  emerging positive flux in this area canceling out the negative flux. 
	  In summary, our findings confirm the scenario that rising magnetic flux 
	  tubes cause the AFS.
\end{abstract}

\section{Introduction}
Arch Filament Systems (AFS) have been described by \citet{cit:howardharvey}, \citet{cit:martres66} and \citet{cit:bruzek67, cit:bruzek69}. They appear preferentially in young sunspot groups where magnetic flux is still emerging. The single fibrils of AFSs (visible in chromospheric lines) connect opposite magnetic polarities crossing the `Polarity Inversion Line' (PIL). Moderate upflows are found in the central parts of the fibrils, while strong downflows are detected towards the ends of the fibrils. This scenario is explained by rising flux tubes that transport matter to higher layers in their more or less horizontal parts, and this matter then streams down towards the footpoints of the fibrils \citep{cit:bruzek69}. Recently, AFS have been studied with higher spatial resolution and with 
spectro-polarimetry by \citet{cit:saminat}, \citet{cit:spadaro}, \citet{cit:lagg07}, \citet{cit:xu10}, \citet{cit:vargas}, \citet{cit:grigoreva}, and \citet{cit:ma}. 
A study of the magnetic field in deep photospheric layers and the corresponding velocity field of the same active region 
considered in the present work has been presented by \citet{cit:balthasar16}. Similar observations of different active regions are 
studied by \citet{cit:sergio} and \citet{cit:verma}.
Nevertheless it remains an open question how the different layers of the solar atmosphere are connected and where exactly to locate 
the footpoints of the AFS fibrils in the photosphere.
In this work we investigate the upper photosphere using a silicon line.

\section{Observations and Data Reduction}

\articlefigure{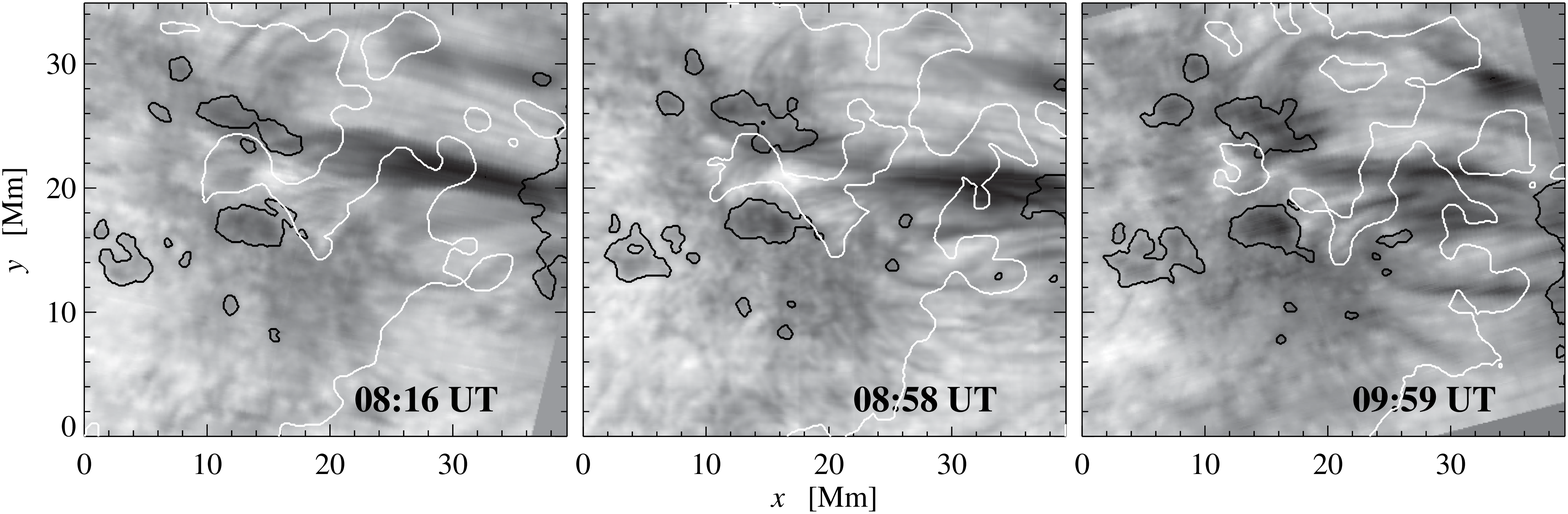}{fig_inthe} 
{Line-core intensities of the `red' component of the helium line at 1083.0~nm. 
Black contours outline the pores and a part of the penumbra at the right edge, and the white line indicates the PIL derived from the silicon line.}

The following part of active region NOAA\,12353 was observed on May 24, 2015 with the GREGOR solar telescope in Tenerife \citep{cit:gregor}. The images were stabilized with an adaptive optics system \citep[GAOS,][]{cit:gaos}.
Spectropolarimetric data were obtained with the GREGOR Infrared Spectrograph \citep[GRIS,][]{cit:gris} which is equipped with the former Tenerife Infrared Polarimeter II \citep{cit:tip2}. The observed spectral range included the spectral line Si~{\sc i} 1082.7~nm, the triplet He~{\sc i} 1083.0~nm, and the line Ca~{\sc i} 1083.9~nm. 
The basic data reduction steps follow the description of \citet{cit:mcv},
and the polarimetric calibration was performed with the GREGOR polarimetric calibration unit \citep{cit:gpu}.
Within two hours, we took five scans across the region, but because of the page limit we show only three of them here. 
More details about these observations are given by \citet{cit:balthasar16}. 
Figure\,\ref{fig_inthe} displays the line-core intensities in 
the so-called `red' component of the helium triplet formed in the upper chromosphere. The fibrils of the AFS appear as dark stripes connecting the main spot (outside the scan region except for the outer penumbra) with the area of the following pores.

The profiles of the photospheric lines (Si~{\sc i} 1082.7~nm and Ca~{\sc i} 1083.9~nm) were inverted using the code 
`Stokes Inversion based on Response functions' \citep[SIR,][]{cit:sir}. For details of the inversion of the calcium line, we refer to \citet{cit:balthasar16}. 
Because the silicon line is sensitive over a wider height range, we set five nodes for temperature in the second and third 
iteration cycle (instead of three, as used for the calcium line).
The magnetic field parameters were kept height independent.
The magnetic azimuth ambiguity was resolved in the same way as for the calcium line, and finally, all images have been derotated numerically by applying the method described in the appendix of \cite{cit:balthasar16}.

\articlefigure{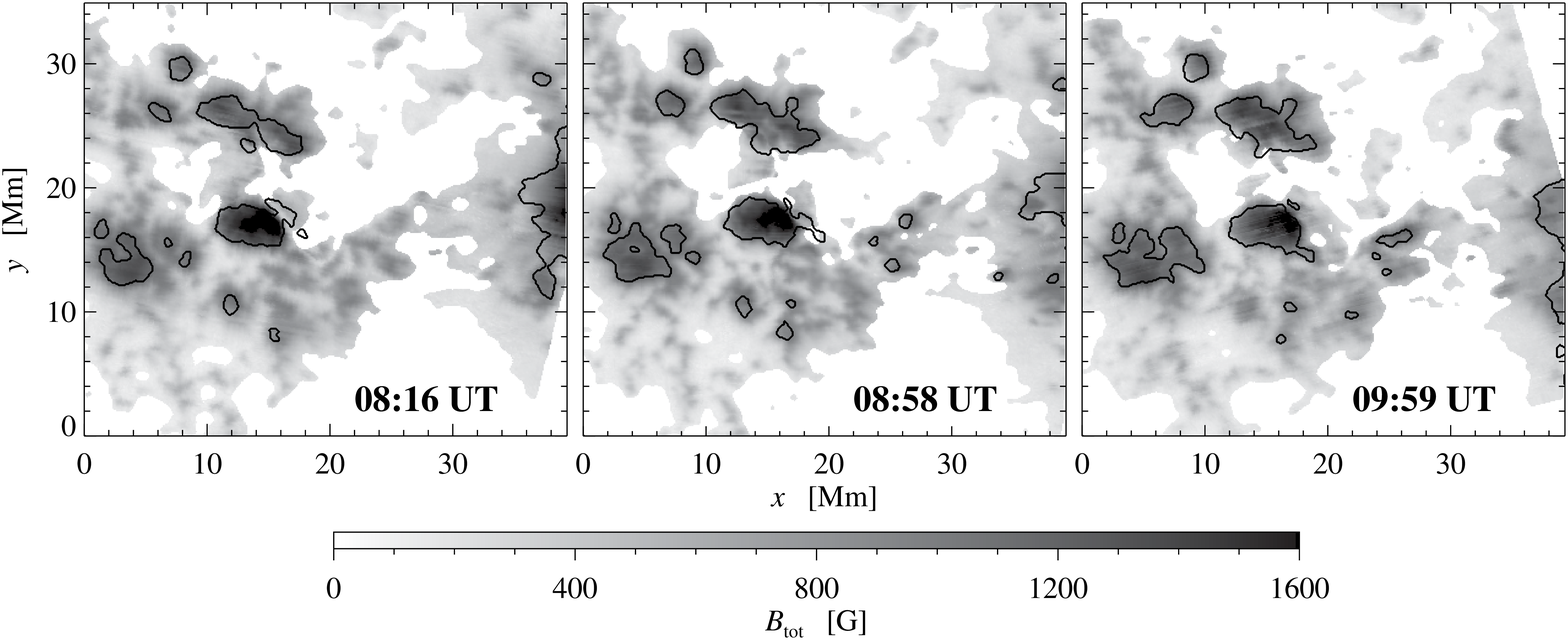}{fig_btot} 
{Total magnetic field strength derived from the silicon line. 
Black contours outline the dark photospheric structures. Values from areas with insignificant polarization signal are suppressed 
and shown in white.}

The Doppler velocities for the silicon line were also obtained from Stokes-$I$ profiles with a Fourier phase method described by \citet{cit:velofft}. 
This method yields the position of the line profile as a whole.
The positions of nearby telluric water vapor lines were used to compensate for shifts caused by the spectrograph. Their mean value in the field-of-view was assumed to be zero.

\section{Results}
The total magnetic field strength is displayed in Fig.~\ref{fig_btot}. The maximum values derived from the silicon line are much lower 
than those from the calcium line \citep[see][]{cit:balthasar16}. Here we encounter only values up to 1600~G, except for the central pore ($x$\,=\,14~Mm, $y$\,=\,18~Mm) where the maximum value is 1920~G. 
This difference with respect to the calcium line is caused by the distinct formation heights of the two spectral lines, the silicon line forming higher than the calcium one. 
Outside the pores, the polarization signal from the silicon line is stronger, and we detect a magnetic field in a more extended spatial range than for the lower photosphere. 
In Fig.~\ref{fig_gamm} we show the inclination of the magnetic field in the local reference frame. Values below 90$^\circ$ indicate a positive polarity which we detect for the pores. The main spot on the right has a negative polarity. Between the pores we find an area 
with negative polarity. 
Determined from the silicon line, this area was connected with the main spot, and the fibrils of the AFS lay above this connection, 
with some of them crossing the winding PIL (see Fig.~\ref{fig_inthe}) three times. 
Such a connection of negative polarity was not detectable for the calcium line \citep[see][]{cit:balthasar16}. Next to the pores, 
this area of negative polarity decreased in size during our observations, as already found for the calcium line \citep[see][]{cit:balthasar16}.

\articlefigure{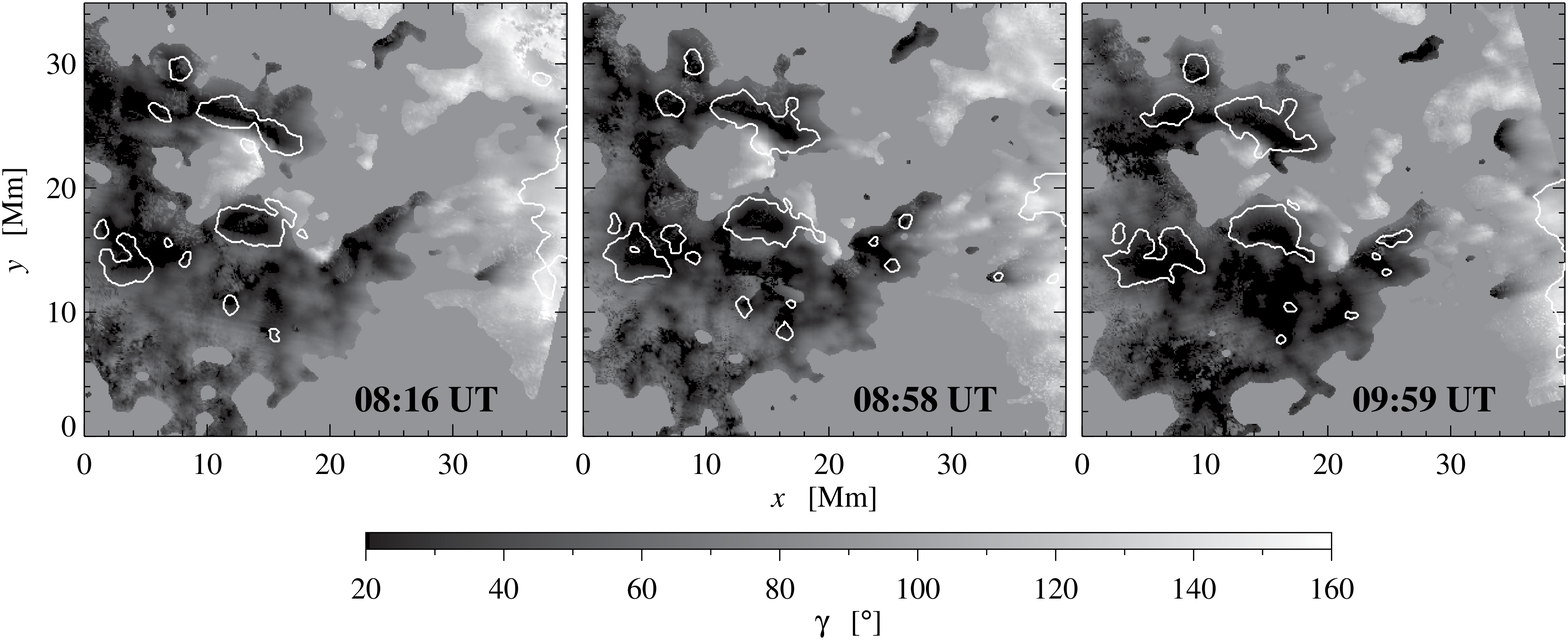}{fig_gamm}
{Inclination of the magnetic field derived from the silicon line. White contours outline the dark photospheric structures. 
Values from areas with insignificant polarization signal are suppressed and shown in medium gray.}

The Doppler velocities obtained from the silicon line are displayed in Fig.~\ref{fig_velo}. 
They exhibit a granular pattern outside the pores, although the contrast is smaller than for the calcium line 
(due to the higher formation layers). In the region below the AFS fibrils, 
patches with the same direction of the velocities are somewhat larger than in the quiet Sun, 
and there is a tendency that strong blueshifts occur preferentially close to the PIL on the side with the positive magnetic 
polarity of the pores. In the penumbra, we mainly detect redshifts, although one would expect blueshifts according to the photospheric Evershed effect, since the disk center is towards the left in the images. However, the silicon line is formed in the upper photosphere, 
and therefore it is less sensitive to the Evershed effect which is much stronger in the deep photosphere. In the left panel at 08:16 UT, there is one patch at the boundary of the penumbra of the main spot with a very high redshift of 4~km~s$^{-1}$ that is likely related to the right end of a dark AFS-fibril. The maximum velocity occurs between the position of the maximum seen in the helium line and that of 
the calcium line \citep[see][]{cit:balthasar16}. Downflows between the two major pores reach 2.4~km~s$^{-1}$, structured in several small patches. In one of these patches indicated by arrows in Fig.~\ref{fig_velo}, the downflow turns into an upflow of 0.8~km~s$^{-1}$.

\articlefigure{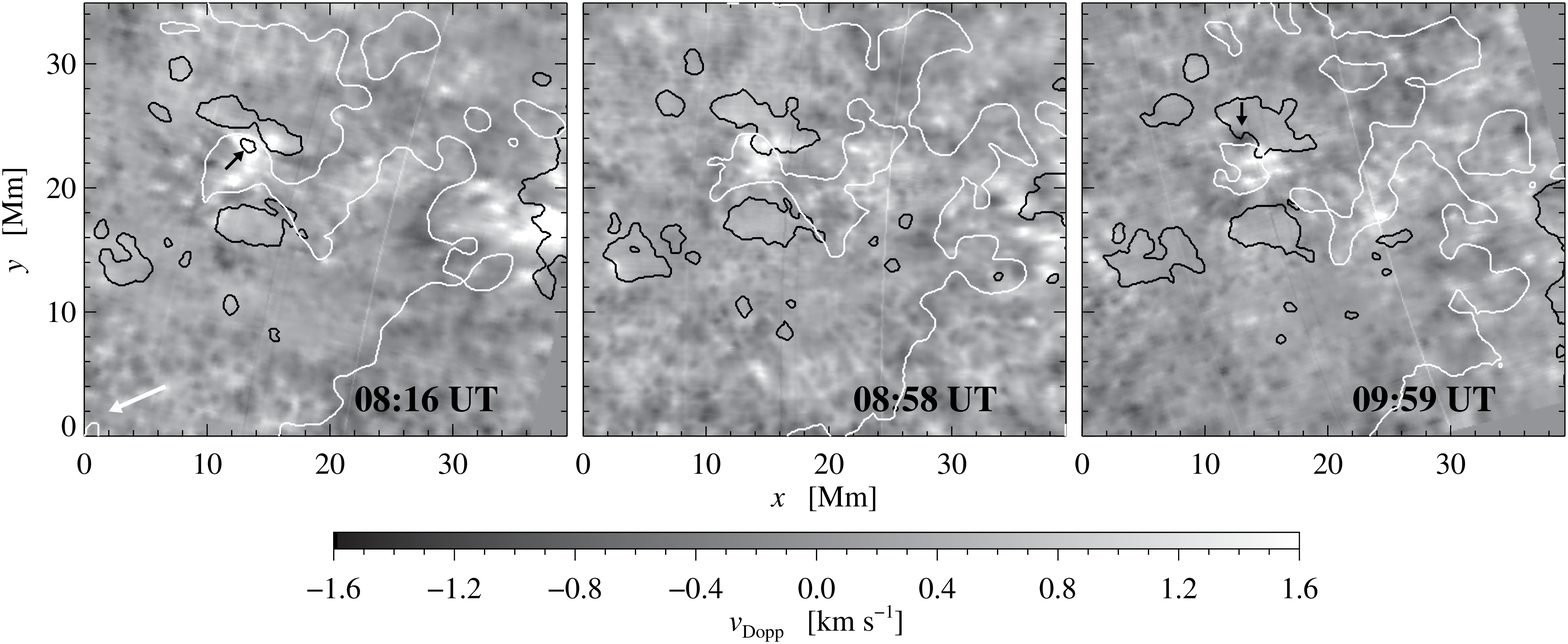}{fig_velo}
{Doppler velocities derived from the silicon line. 
Black contours outline the dark photospheric structures, and the white contours indicate the PIL.  
The display is limited to the range $\pm$1.6~km~s$^{-1}$. The small black arrows point to a small patch where the sign of the Doppler shifts changed from the first to the last scan. The white arrow points to the disk center.}

\section{Discussion and Conclusions}
It is difficult to identify photospheric structures that are directly related to the chromospheric structures as seen in the helium line. This holds especially for the footpoints of the fibrils of the AFS.
This problem is also pointed out by \citet{cit:schad}. While the silicon line probes the upper photosphere up to 500~km the helium line 
is formed in much higher layers, probably even higher than 1000~km \citep[see e.g.,][]{cit:kmws}. 
\citet{cit:merenda} found that loops in a region with emerging flux reach up to more than 5000~km, and they obtained the best fit 
for a top height of 6300~km. For a reconstruction of the magnetic field, \citet{cit:xu10} assume a loop height of 3000\,km. With the present data, we cannot close this gap.

In the scenario of rising flux tubes, one would expect the highest velocities close to the very ends of the fibrils.
However, during our first scan we find the highest Doppler velocities from the silicon line just beside the visible ends of the fibrils above the spot. In addition, looking at the positions of highest velocities deduced from the calcium and the helium lines \citep{cit:balthasar16}, we find that the highest velocities from the helium line are further to the West than the
maximum from the calcium line, and the maximum from the silicon line falls in between. From a simple rising loop scenario one 
would expect the opposite, but the strong magnetic field of the spot itself is near, and in higher layers where the 
influence from the sunspot is less, the loop expands for more than the distance of the photospheric footpoints. 

Between the pores, we encounter magnetic flux of opposite polarity compared to the positive polarity of the pores. 
The area of this negative flux decreases during the period of our observations, which we explain by rising magnetic loops.
The loops appear first close to the PIL and have the same magnetic orientation as the whole sunspot group. 
While rising, the small positive footpoints move towards the area between the pores, and field lines reconnect. 
In an alternative scenario the negative flux is return flux related to the pores and is already connected below 
the solar surface with the emerging loops. When the loops rise, the magnetic field lines straighten, and the negative polarity 
is no longer visible.

Our results confirm the scenario of rising flux tubes creating the fibrils of the AFS, as it was proposed by \citet{cit:bruzek67, cit:bruzek69} and confirmed by \citet{cit:xu10}. Nevertheless, the silicon data show that the real magnetic configuration is more 
complex than this simple picture. The PIL is winded, and the fibrils of the AFS cross the PIL three times. 
Our work to invert the helium line is still in progress, and we shall report 
on the results in a future article.

\acknowledgments 
The 1.5-meter GREGOR solar telescope was built by a German 
consortium under the leadership of the Kiepenheuer-Institut f\"ur Sonnenphysik 
in Freiburg (KIS) with the Leibniz-Institut f\"ur Astrophysik Potsdam (AIP), 
the Institut f\"ur Astrophysik G\"ottingen (IAG), the Max-Planck-Institut f\"ur 
Sonnensystemforschung in G\"ottingen (MPS), and the Instituto de 
Astrof{\'\i}sica de Canarias (IAC), and with contributions by the Astronomical 
Institute of the Academy of Sciences of the Czech Republic (ASCR). SJGM is 
grateful for financial support from the Leibniz Graduate School for 
Quantitative Spectroscopy in Astrophysics, a joint project of AIP and the 
Institute of Physics and Astronomy of the University of Potsdam. 
This work was supported by a program  of the Deutscher Akademischer 
Austauschdienst (DAAD) and the Slovak Academy of Sciences for project related 
personnel exchange (project No.~57065721). 
This work received additional support by project VEGA 2/0004/16. 
The GREGOR observations were obtained within the SOLARNET Transnational Access 
and Service (TAS) program, which is supported by the European Commission's FP7 
Capacities Program under grant agreement No.~312495.

\end{document}